\documentclass[aps,tightline,nofootinbib,superscriptaddress]{revtex4}
\usepackage{graphicx}
\usepackage{amssymb}
\usepackage{amsmath,bm}
\usepackage{afterpage}
\usepackage{float}
\usepackage{epsfig}
\usepackage{subfigure}
\usepackage{ulem}

\begin{document}

\title{RGE of the Higgs mass in the context of the SM}

\author{Ligong Bian}
 \email{lgb@mail.nankai.edu.cn}

\affiliation{%
 School of Physics, Nankai University, Tianjin 300071, China
}

\date{\today}

\begin{abstract}

Renormalization group equation (RGE) of the Higgs mass up to one-loop order is
derived in $R_\xi$ gauge with dimensional regularization method (DREG) base on MS
(or $\overline{\rm{MS}}$) scheme.
Scale-dependent properties of the Higgs mass up to two-loop
level are investigated. The result gives the positive
 Higgs mass up to energy scale around the Planck scale.
While following the vacuum stability argument, the Higgs mass
need to be negative above the energy scale $10^{10}$ GeV for $M_{t}=173.3$ GeV.
The conflict is induced by the renormalization of vacuum energy value(VEV).

\end{abstract}


\maketitle

\section{Introduction}

For the appearance of the hierarchy problem (or the naturalness problem), it is not very
easy to do ordinary perturbation theory calculation, i.e., treat divergences with
regularization method and choose renormalization prescript to
calculate renormalization group equations (RGEs).
The famous naturalness problem was studied in 't Hooft-Feynman gauge completely three
decades before~\cite{Veltman:1980mj}. In order to manifest quadratic divergences,
dimensional regularization (DREG) is used and the poles at $d=2$ are kept,
and the correspondence $1/(1-d/2)\rightarrow\Lambda^{2}/(4\pi)$ was adopted.
The quadratic
divergences lives on the complex two dimensional
plane in the sense of~\cite{Veltman:1980mj}, then we can proceed renormalization of the
Higgs mass term at $d=4$ safely. Indeed, DREG preserves the property that
the poles at $d=2$ and $d=4$ can be renormalized independently~\cite{'tHooft:1972fi,Leibbrandt:1975dj},
poles at $d=2$ won't shows up when the dimensionality is compacted to 4,
which allow us to study the RGEs of parameters of SM at $d=4$.

The RGE of the Higgs mass is derived in section.~\ref{sec:RGS}. We choose DREG to treat
divergences and use MS (or $\overline{\rm{MS}}$) scheme as the renormalization scheme.
In order to make our results more universality, all calculations in this work are
proceeded in $R_{\xi}$ gauge. The renormalization constant of the Higgs mass, wherein poles at
$d=2$ been kept, is also given in this work to derive the formula which expresses the hierarchy problem.
After RGE of the Higgs mass been derived, we investigated scale-dependent properties of the
Higgs mass and that of the Higgs quartic coupling.
The scale-dependent properties of the Higgs mass and the Higgs quartic coupling are studied in
section.~\ref{sec:NumericalAnalysis}.
The section.~\ref{sec:Higgsmass} is devoted to
discussions and conculsions.

\section{Renormalization of the Higgs mass}
\label{sec:RGS}

Direct physical quantities couldn't be gauge dependent. In order to extract some useful
physical consequences, we need to explore RGEs of parameters of renormalizable theory in MS
(or $\overline{\rm{MS}}$) scheme,
which has the remarkable property that in this scheme beta functions ($\beta$) and anomalous
dimension of the mass parameter ($\gamma_m$) are gauge-independent. The argument on this viewpoint is
given by~\cite{Caswell:1974cj,Gro:76}.
In other renormalization schemes, the renormalization coupling constant ($Z_{g,m}$) for
coupling or mass is, in general, gauge dependent.
This is caused by appearance of the finite terms which dependent on g and $\xi$, in
addition to the terms given in $Z_{g,m}$, with
\begin{eqnarray}\label{eq:renor_cons}
¡¡Z_{g,m}(g,\xi)=1+\Sigma_{\nu=1}\frac{Z^{(\nu)}_{g,m}(g,\xi)}{\varepsilon^{\nu}},
\end{eqnarray}
where $g,\xi$ are renormalized couplings and gauge parameter, respectively.
We should note that there is no explicit scale parameter $\mu$-dependent shown in Eq.~(\ref{eq:renor_cons}),
so RGEs of couplings (beta functions) and anomalous dimension of mass operator $\gamma_{m}$
in MS (or $\overline{\rm{MS}}$) scheme( which are functions of $Z_{g,m}$) carry no explicit
$\mu$-dependent. That's way MS (or $\overline{\rm{MS}}$) scheme is always referred as the mass-independent
renormalization scheme. This property of MS (or $\overline{\rm{MS}}$) scheme makes it very
easy to solve RGEs. The above argument construct the motivation to derive RGE of the Higgs mass
in MS (or $\overline{\rm{MS}}$) scheme in this work.

\subsection{Renormalization procedure and the VEV}

Consider the Higgs mass of the SM as one physical quantity,
then the two-point connected Green function of the Higgs field
need to be gauge invariant. When we study the perturbative correction
of the Higgs mass of the SM, one should take into account not only
standard loop corrections (the two point 1PI self-energies),
but also corrections to the definition of VEV
via minimal of the Higgs potential. The loop corrections to the definition of VEV,
entering through the so called tadpole 1PI (one point 1PI truncated Green function),
cause VEV shift.
The VEV shift induced by tadpole 1PI values much in the mass renormalization,
which induced the 1PR two-point self-energy of the Higgs field~\cite{Ma:1992bt},
with which can we get gauge invariant mass correction~\cite{Weinberg:1973ua,Fleischer:1980ub}.
This kind of renormalization procedure have also been
adopted in the case of fermion mass renormalization~\cite{Hempfling:1994ar},
and for the renormalization of other mass terms in the SM involving VEV,
the same procedure need to be adopted in order that we get gauge invariant
masses.

One more thing, the tadpole 1PI is related with the Higgs potential in the sense of
~\cite{Coleman:1973jx}, with the tadpole 1PI we can arrive at the
Higgs potential straightforwardly (see Eq.
(3.20) and (3.21) of the paper~\cite{Weinberg:1973ua}),
and the relation between the tadpole 1PI and the effective potential
in Landau gauge is given in~\cite{Lee:1974fj},
and that the derivation of the Higgs potential with this
method is technically more easier~\cite{Sher:1988mj}.

\subsection{The Lagarangian and counter-term method}

Relations between renormalized masses and parameters used in this work are
\begin{eqnarray}\label{eq:notation}
 &&m_{H}=\sqrt{2\lambda}~v,\qquad m_{W}=\frac{g_{2}v}{2},\qquad
   m_{Z}=\frac{g_{1}v}{2cos\theta_{W}},\nonumber\\
 &&m_{t}=\frac{g_{t}v}{\sqrt{2}}, ~~\qquad {\rm cos}\theta_W=\frac{g_{2}}{\sqrt{g_{2}^{2}+g_{1}^{2}}}.
 \end{eqnarray}
with $\lambda$, $g_{t}$, $g_{2}$ and $g_{1}$ are scalar quartic coupling, top-quark Yukawa coupling,
SU(2)$_L$ and U(1) gauge couplings, respectively.

After SSB, bare Lagrangian of the Higgs part of the SM is
\begin{eqnarray}
\mathcal{L}_{H}^{0}&=&\frac{1}{2}(\partial_{\mu}H^{0})^{2}-\frac{1}{2}(m^{0}_{H})^{2}(H^{0})^{2}
-\lambda^{0}
v^{0}(H^{0})^{3}-\frac{1}{4}\lambda^{0} (H^{0})^{4} \nonumber\\
&& + \mathrm{const.},
\end{eqnarray}
where superscripts ``0'' on mass, couplings and the Higgs field are used to denote that these
parameters are bare parameters, parameters which do not have superscripts represent
renormalized parameters of the SM, and $m^{0}_{H}=\sqrt{2\lambda^{0}}v^{0}$ have been sat
when the above equation is written.
Let us first introduce four renormalization constants to relate bare and renormalized parameters,
\begin{eqnarray}\label{eq:rgec}
\lambda^{0}=Z^{-2}_{H}Z_{1}\lambda,\quad (m^{0}_{H})^{2}=Z^{-1}_{H}Z_{0}(m_{H})^{2},\quad H^{0}=Z^{1/2}_{H}H.
\end{eqnarray}
Then, the relation between bare and renormalized vacuum energy (VEV) is given
by $v^{0}=Z^{-1/2}_{1}Z^{1/2}_{H}Z^{1/2}_{0}v$,
the bare Lagrangian of the Higgs part $\mathcal{L}_{H}^{0}$ can be separated to renormalized part
$\mathcal{L}_{H}$ and the
counter-term Lagrangian $\mathcal{L}^{ct}_{H}$, with $\mathcal{L}_{H}$ precisely equal
to $\mathcal{L}_{H}^{0}$
if bare parameters in $\mathcal{L}_{H}^{0}$ are replaced by the renormalized ones, and
\begin{eqnarray}\label{eq:lagarangian_ct}
\mathcal{L}^{ct}_{H}=\frac{1}{2}(Z_{H}-1)
(\partial_{\mu}H)^{2}-\frac{1}{2}(Z_{0}-1)(m_{H})^{2}
H^{2}-(Z^{1/2}_{1}Z^{1/2}_{0}-1)\lambda vH^{3}-\frac{1}{4}(Z_{1}-1)\lambda H^{4}
 + \mathrm{const.},
\end{eqnarray}
with renormalization constant for the Higgs mass $Z_{m}=Z^{-1}_{H}Z_{0}$ and scalar quartic
coupling renormalization constant $Z_{\lambda}=Z^{-2}_{H}Z_{1}$,
then one arbitrary mass parameter $\mu$ can be introduced
through $\lambda^{0}=Z_{\lambda}\lambda\mu^{\varepsilon}$, similarly, scale $\mu$ can also be
introduced through $(g^{0})^{2}_{1,2,t}=Z_{g_{1},g_{2},g_{t}}g^{2}_{1,2,t}
\mu^{\varepsilon}$, with $\varepsilon=4-d$.
Since beta function of scalar quartic coupling ($\beta_{\lambda}$) have been derived decades
before~\cite{Ford:1992mv,Machacek:1983fi,Luo:2002ey}, we will not derive renormalization constant
$Z_{1}$ anymore, but use $\beta_{\lambda}$ directly.

\begin{figure}
  \centering
  \includegraphics[width=0.5\linewidth]{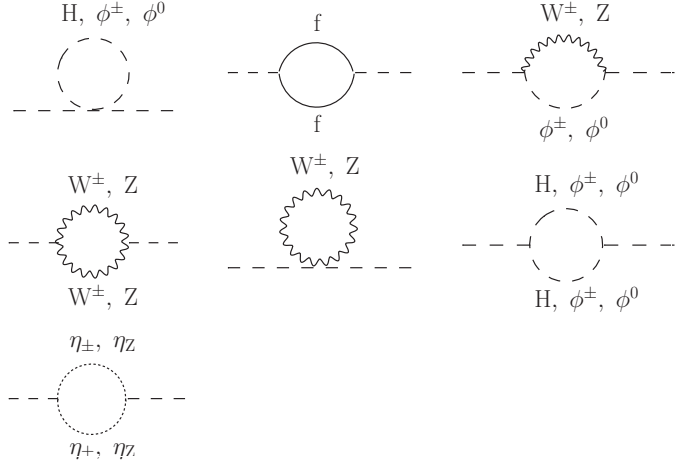}\\
  \caption{One-loop 1PI self-energy corrections to the Higgs mass }\label{Fig:Higgs1PI}
\end{figure}

\begin{figure}
  \includegraphics[width=0.5\linewidth]{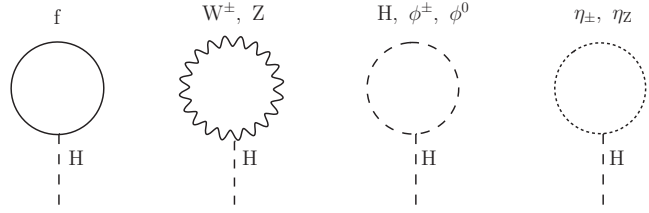}\\
  \caption{Tadpole feynman diagram with one external Higgs field in the SM }\label{fig:tad1loop}
\end{figure}

In MS (or $\overline{\rm{MS}}$) scheme, divergent terms in self-energy of the Higgs field
will be subtracted by renormalization constant. Two-point self-energy of the Higgs field comes
from 1PI self-energy and tadpole
contributions,
\begin{eqnarray}
\Sigma_{H}(p^{2})=\Sigma_{H}^{\rm{1PI}}(p^{2})+\Sigma_{H}^{T}(p^{2})
\end{eqnarray}
with feynman diagrams contribute to 1PI self-energy shown in Fig.~\ref{Fig:Higgs1PI},
and the tadpole diagrams contribution to the Higgs self-energy is
\begin{equation}
\Sigma_{H}^{T}=-i\frac{3(m_{H})^{2}}{v}\frac{i}{-(m_{H})^{2}}T,
\end{equation}
where $\frac{i}{-(m_{H})^{2}}$ is the propagator of the Higgs boson carrying $zero$ momentum,
the Higgs three-point vertex is $\frac{-i3(m_{H})^{2}}{v}$, with ``T'' represent feynman diagrams
contributions shown in Fig.~\ref{fig:tad1loop},

Up to one-loop level, the counter-term method requires
$\Sigma_{H}(p^{2})+i(Z_{H}-1)p^{2}-i(Z_{0}-1)m^{2}_{H}=0$, combined with the second formula given
in Eq.~(\ref{eq:rgec}) and the relation $Z_{m}=Z^{-1}_{H}Z_{0}$, we can derive renormalization
constant of the Higgs mass ($Z_{m}$).

\subsection{Renormalization coupling constant $Z_{m}$ and anomalous dimension of the Higgs mass $\gamma_{m_{H}}$}

Scalar momentum integral which can give rise to quadratic divergences involved in self-energy
calculations are shown below and other momentum integrals are listed in section.~\ref{sec:MI}
\begin{eqnarray}\label{eq:div}
\int\frac{d^{d}k}{(2\pi)^{d}}\frac{1}{k^{2}-m^{2}}
=-i\frac{1}{4\pi}\frac{1}{1-d/2}+i\frac{m^{2}}{2-d/2}
\end{eqnarray}
where two poles $1/(1-d/2)$ and $1/(2-d/2)$ are all kept, while when one compute the momentum
integration shown in Eq.~(\ref{eq:div}), one can expand results around $d=2$
or $d=4$, which give two different pole,
when dimension d continued to ``4'', corresponding to four dimensionality physics,
then there is no pole at $d=2$, which give
quadratic divergences on the complex two dimensional plane in the sense of~\cite{Veltman:1980mj}.

Proceeding calculations in DREG and keeping poles $1/(2-d/2)$, renormalization constant of the
Higgs field can be calculated,
\begin{eqnarray}\label{eq:ZH}
Z_{H}=1-\frac{1}{(4\pi)^{2}}\frac{1}{2-d/2}\big[\frac{g^{2}_{2}
(\xi-3)}{2}+\frac{g^{2}_{1}+g^{2}_{2}}{4}(\xi-3)+3g^{2}_{t}\big],
\end{eqnarray}
and renormalization constant is derived as:
\begin{eqnarray}\label{eq:Z0}
Z_{0}=1+\frac{1}{(4\pi)^{2}}\frac{1}{2-d/2}
\left(6\lambda-\frac{3\xi}{4}g^{2}_{2}-\frac{\xi}{4}g^{2}_{1}\right).
\end{eqnarray}
Thus, renormalization constant of the Higgs mass on the basis of MS (or $\overline{\rm{MS}}$)
scheme is calculated to be
\begin{eqnarray}\label{eq:mconst}
Z_{m}&=&Z^{-1}_{H}Z_{0}\nonumber\\
&=&1+\frac{1}{(4\pi)^2}\frac{2}{4-d}\left(6\lambda+3g^{2}_{t}-\frac{9}{4}g^{2}_{2}-\frac{3}{4}g^{2}_{1}\right),
\end{eqnarray}
it is obviously that $Z_{m}$ is gauge independent.

RGE of the Higgs mass in MS (or $\overline{\rm{MS}}$) can be given by
\begin{eqnarray}\label{eq:gamH}
\mu\frac{dm_{H}}{d\mu}&=&-m_{H}\lim_{\varepsilon\rightarrow 0}\gamma_{m_{H}}(m_{H}(\mu),\varepsilon),
\end{eqnarray}
with
\begin{eqnarray}
\gamma_{m_{H}}(m_{H}(\mu),\varepsilon)&=&\frac{\mu}{2Z_{m_{H}}}
\frac{dZ_{m}}{d\mu}\nonumber\\
&=&\frac{\mu}{2Z_{m}}\left(\frac{\partial Z_{m}}{\partial \lambda}\beta(\lambda(\mu),\varepsilon)
+\frac{\partial Z_{m}}{\partial g_{t}}\beta(g_{t}(\mu),\varepsilon)+\frac{\partial Z_{m}}{\partial g_{1}}\beta(g_{1}(\mu),\varepsilon)+\frac{\partial Z_{m}}{\partial g_{2}}\beta(g_{2}(\mu),\varepsilon)\right)
\end{eqnarray}
represent anomalous dimension of the Higgs mass term, wherein
\begin{eqnarray}\label{eq:beta_d}
\beta(\lambda(\mu),\varepsilon)&=&-\varepsilon\lambda+
\beta_{\lambda}=\mu\frac{d\lambda(\mu)}{d\mu},\\
\beta(g_{t}(\mu),\varepsilon)&=&-\frac{\varepsilon}{2} g_{t}+\beta_{g_{t}}=\mu\frac{dg_{t}(\mu)}{d\mu},\\
\beta(g_{1}(\mu),\varepsilon)&=&-\frac{\varepsilon}{2} g_{1}+\beta_{g_{1}}=\mu\frac{dg_{1}(\mu)}{d\mu},\\
\beta(g_{2}(\mu),\varepsilon)&=&-\frac{\varepsilon}{2} g_{2}+\beta_{g_{2}}=\mu\frac{dg_{2}(\mu)}{d\mu},
\end{eqnarray}
are RGEs of $\lambda,~g_{t},~g_{1},~g_{2}$ (beta functions), the additional factor ``$\frac{1}{2}$''
before $g_{t,1,2}$ in the last
three equations compare with the first equation is caused by the introduction approach of $\mu$, as
described below
Eq.~(\ref{eq:lagarangian_ct}),
and they reduce to beta functions: $\beta_{\lambda},~\beta_{g_{t}},~\beta_{g_{1}}~,\beta_{g_{2}}$
in four dimensional space-time.

Since the anomalous dimension of the Higgs mass ($\gamma_{m_{H}}$) is function of $Z_{m}$, it must
be gauge independent. While in other renormalization schemes the renormalization constant for
$m_{H}$ is, in general, gauge dependent. This is caused by appearance of the finite terms
in addition to the terms given in $Z_{m}$ at the right side of Eq.~(\ref{eq:mconst})
in other schemes.

RGE of the Higgs mass in $\overline{\rm{MS}}$ scheme in four dimensional space-time is given by
 \begin{align}\label{eq:rgem}
 \mu\frac{\mathrm{d}m_{H}}{\mathrm{d}\mu}
 =-m_{H}\gamma_{m_{H}},
 \end{align}
with anomalous dimension of the Higgs mass up to one-loop level
 \begin{eqnarray}\label{eq:gammH}
 \gamma^{(1)}_{m_{H}}=-\frac{1}{16\pi^{2}}\left(6\lambda+3g_t^2-
 \frac{9}{4}g_2^2-\frac{3}{4}g_1^2\right),
 \end{eqnarray}
here one need to note that others Yukawa couplings except of that of top quark are all dropped
for small contributions. We found that RGE of $m_{H}$ derived by us, i.e.,
Eq.~(\ref{eq:rgem}) has the same formula with the RGE of the Higgs mass parameter $m$
~\cite{Luo:2002ey}, which is inferred from the results in~\cite{Machacek:1983fi} through introducing "dummy" real scalar field.
If the $m$ there is replaced by $m_{H}$ and with the notations used there been translated to ours, one can get Eq.~(\ref{eq:rgem}).
Thus both two methods can give the same
gauge invariant RGE of the Higgs mass.

For the two-loop case, the gauge invariant property should be kept also,
and we can safely infer that the anomalous dimension at two-loop order with our method explored
in this paper need to have the same form as that given by~\cite{Luo:2002ey}.
For complexity, we will just leave the derivation
of the anomalous dimension at two-loop order for further study, and use the result
derived in~\cite{Luo:2002ey}, which is specialized with the SM couplings in~\cite{Holthausen:2011aa}:
 \begin{eqnarray}\label{eq:gammH2}
 \gamma^{(2)}_{m_{H}}&=&-\frac{1}{(16\pi^{2})^{2}}
 \bigg(-30\lambda^{2}
 -36\lambda g^{2}_{t}+12\lambda(3g^{2}_{2}+g^{2}_{1})-\frac{27}{4}
 g^{4}_{t}+20g^{2}_{3}g^{2}_{t}+\frac{45}{8}g^{2}_{2}g^{2}_{t}
 \nonumber\\
 &&+\frac{85}{24}
 g^{2}_{1}g^{2}_{t}-\frac{145}{32}g^{4}_{2}+\frac{15}{16}g^{2}_{1}
 g^{2}_{2}
 +\frac{157}{96}g^{4}_{1}\bigg),
 \end{eqnarray}
and it's validity will be justified in the next section through investigating
the behaviour of $m_{H}(\mu)$ with respect to $\mu$.

While when we proceed calculations, we keep all poles at $d=2$ and $d=4$ for completeness,
renormalization constant for the Higgs field $Z_{H}$ will be the same as Eq.~(\ref{eq:ZH}),
while the renormalization constant including only contributions from terms that represent poles at
$d=2$ (denoted as $Z'_{0}$ to differentiate from the renormalization constant obtained when
the pole is achieved at $d=4$($Z_{0}$)) is
\begin{eqnarray}
Z'_{0}=1+\frac{2}{(4\pi)m^{2}_{H}}\frac{1}{1-d/2}\big[6\lambda-
\frac{3}{2}Tr[I]g^{2}_{t}
+(g^{\mu}_{\mu}-1)(\frac{3g^{2}_{2}}{4}
+\frac{g^{2}_{1}}{4})\big]+\frac{1}{(4\pi)^{2}}\frac{1}{2-d/2}
\left(6\lambda-\frac{3\xi}{4}g^{2}_{2}-\frac{\xi}{4}g^{2}_{1}\right).
\end{eqnarray}
The renormalization constant of the Higgs mass on the complex two dimensional plane (denoted as $Z'_{m}$
to differentiate from the renormalization constant obtained when
the pole is achieved at $d=4$ ($Z_{m}$)) is
calculated to be:
\begin{eqnarray}\label{eq:Zm'}
Z'_{m}&=&Z^{-1}_{H}Z'_{0}\nonumber\\
&=&1+\frac{2}{(4\pi)(m_{H})^{2}}\frac{1}{1-d/2}\big[6\lambda-
\frac{3}{2}Tr[I]g^{2}_{t}
+(g^{\mu}_{\mu}-1)(\frac{3g^{2}_{2}}{4}+\frac{g^{2}_{1}}{4})\big] +\frac{1}{(4\pi)^2}
\frac{2}{4-d}\left(6\lambda+3g^{2}_{t}
-\frac{9}{4}g^{2}_{2}-\frac{3}{4}g^{2}_{1}\right).
\end{eqnarray}

From Eq.~(\ref{eq:Zm'}), we found that if we take replacement
$1/(1-d/2)\rightarrow\Lambda^{2}/(4\pi)$
(which can be obtained when one compare Eq.~(\ref{eq:div}) with the same integral that
calculated with naive cut-off method in four dimensional space-time), and consider
poles at $d=2$ alone, from the fact that dimension d couldn't compacted to ``2'' and
``4'' at the same time in the sense of Eq.~(\ref{eq:div}), then the hierarchy problem
can be expressed by
\begin{equation}\label{eq:DRED}
(m^{0}_{H})^{2}=(m_{H})^{2}+\frac{2\Lambda^{2}}{(4\pi)^{2}v^{2}}
\big[3(m_{H})^{2}-12(m_{t})^{2}
+6(m_{W})^{2}+3(m_{Z})^{2}\big],
\end{equation}
To get this formula, Tr[I]=$g^{\mu}_{\mu}$=4 and Eq.~(\ref{eq:notation}) need to be
adopt in the second term at right side of Eq.~(\ref{eq:Zm'}). Therefore, we derived the expression
of naturalness, i.e., Eq.~(\ref{eq:DRED}), with the quadratic divergences manifested as pole
at $d=2$ on the complex two dimensional plane~\cite{Veltman:1980mj}. And the naturalness problem
is manifestly gauge independent in our remormalization procedure, as shown in Eq.~(\ref{eq:DRED}).

\section{Scale-dependent property of the Higgs mass}
\label{sec:NumericalAnalysis}

Since only when all RGEs in the SM been studied together, can we investigate physics
in the system of the SM, i.e., all electro-weak couplings ($\lambda,~g_{1},~g_{2},~g_{t}$)
and QCD couplings $g_{3}$ and the Higgs mass together compose the whole physical system of the SM.
We explore the $\mu$-dependent property of the Higgs mass with all RGEs of all couplings of
the SM been considered for the first time.

The beta function for a generic coupling \(X\) needed is given as:
\begin{align}\label{eq:betafunctions}
\mu\frac{\mathrm{d}X}{\mathrm{d}\mu}=\frac{\beta_X}{16\pi^2},
\end{align}
with beta functions up to one-loop order~\cite{Einhorn:1992um,Machacek:1983fi}:
\begin{eqnarray}
\beta^{(1)}_{\lambda}&=&\lambda(-9g^2_2-3g^2_1+12g_t^2)+24\lambda^2
+\frac{3}{4}g_2^4+\frac{3}{8}(g_1^2+g_2^2)^2-6g_t^4, \\
\beta^{(1)}_{g_t}&=&\frac{9}{2}g_t^3+g_t\left(-\frac{17}{12}g_1^2
-\frac{9}{4}g_2^2-8g_3^2 \right),\\
\beta^{(1)}_{g_1}&=&\frac{41}{6}g_1^3,\quad \beta_{g_2}=-\frac{19}{6}g_2^3,
\quad \beta^{(1)}_{g_3}=-7g_3^3,
 \end{eqnarray}
and beta functions up to two-loop order~\cite{Machacek:1983fi,Holthausen:2011aa}
 \begin{eqnarray}
 \beta^{(2)}_{\lambda}&=&-312\lambda^3-144 \lambda^2 g_t^2 + 36 \lambda^2 (3 g_2^2 + g_1^2) -
 3 \lambda g_t^4 + \lambda g_t^2
 \left(80 g_3^2 + \frac{45}{2} g_2^2 + \frac{85}{6} g_1^2\right) \nonumber \\
 &&\quad - \frac{73}{8} \lambda g_2^4 + \frac{39}{4} \lambda g_2^2 g_1^2 +
 \frac{629}{24} \lambda g_1^4 + 30 g_t^6 - 32 g_t^4 g_3^2 -
 \frac{8}{3} g_t^4 g_1^2 - \frac{9}{4} g_t^2 g_2^4  \nonumber \\
 &&\quad+ \frac{21}{2} g_t^2 g_2^2 g_1^2 - \frac{19}{4} g_t^2 g_1^4
 + \frac{305}{16} g_2^6 - \frac{289}{48} g_2^4 g_1^2 - \frac{559}{48} g_2^2 g_1^4
 - \frac{379}{48} g_1^6 ,\\
 \beta_{g_t}^{(2)}&=&g_t \left(-12 g_t^4 + g_t^2 \left(\frac{131}{16} g_1^2
 + \frac{225}{16} g_2^2 + 36 g_3^2 - 12 \lambda\right) +\frac{1187}{216} g_1^4\right.\nonumber \\
 &&\quad \left. - \frac{3}{4} g_2^2 g_1^2 + \frac{19}{9} g_1^2 g_3^2
 -\frac{23}{4} g_2^4 + 9 g_2^2 g_3^2 - 108 g_3^4 + 6 \lambda^2\right), \\
 \beta^{(2)}_{g_1}&=&g_1^3 \left(\frac{199}{18} g_1^2 + \frac{9}{2} g_2^2
 + \frac{44}{3} g_3^2 - \frac{17}{6}g_t^2\right), \\
 \beta^{(2)}_{g_2}&=&g_2^3 \left(\frac{3}{2} g_1^2 + \frac{35}{6} g_2^2
 + 12 g_3^2 - \frac{3}{2} g_t^2\right),\\
 \beta^{(2)}_{g_3}&=&g_3^3 \left(\frac{11}{6} g_1^2 + \frac{9}{2} g_2^2
 - 26 g_3^2 -2 \lambda_t^2\right),
 \end{eqnarray}

To study the scale-dependent property of the Higgs mass, we need to solve the RGE of the Higgs mass
 \begin{eqnarray}
  \mu\frac{\mathrm{d}m_{H}}{\mathrm{d}\mu}
 =-m_{H}(\gamma^{(1)}_{m_{H}}+\gamma^{(2)}_{m_{H}}),
 \end{eqnarray}
 with anomalous dimension of the Higgs given by Eq.~(\ref{eq:gammH},\ref{eq:gammH2}),
 together with Eq.~(\ref{eq:betafunctions}) numerically.

\begin{figure}[!htp]
  \centering
  \includegraphics[width=0.45\linewidth]{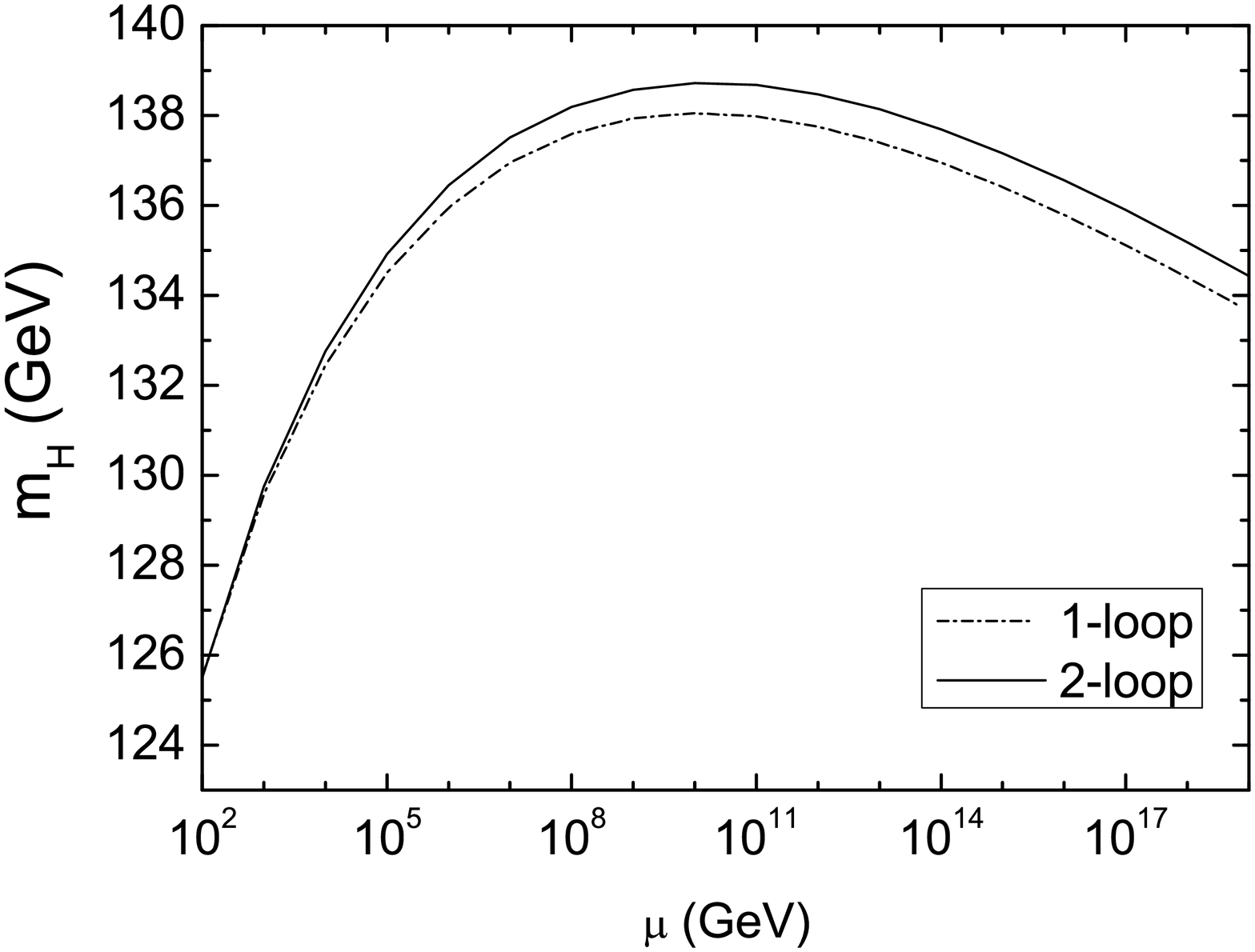}
  \centering
  \centering
   %Requires \usepackage{graphicx}
  \includegraphics[width=0.45\linewidth]{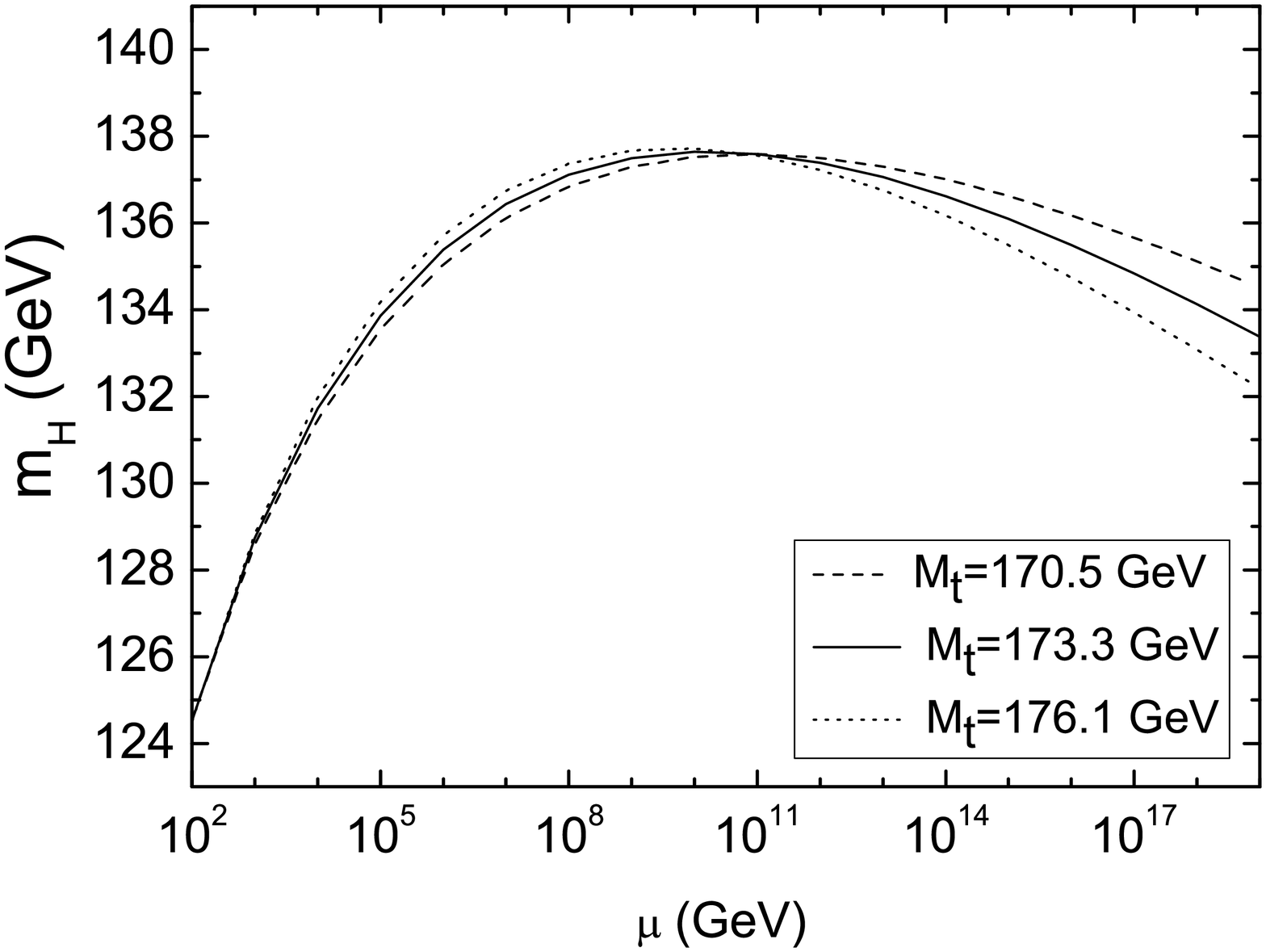}\\
  \caption{Behavior of the Higgs mass with respect to energy scale.
  Left: Scale dependent properties of the Higgs mass up to 1-loop and 2-loop level with $M_{t}=173.3$ GeV; Right: Dependence of $M_{t}$ for the scale dependent property of the Higgs mass up to 2-loop level. }\label{fig:mHmu}
 \end{figure}

 When we solve RGEs of all couplings and the Higgs mass up to two-loop order at the same time,
 boundary conditions of $m_{H}$ is set to $m_{H}=126$ GeV~\cite{Aad:2012tfa} and
  and boundary conditions of $g_{1,2,3,t}$ can be obtained
 as in~\cite{Holthausen:2011aa,Xing:2011aa}.
 When we solve beta functions of all couplings up to two-loop order~\cite{Luo:2002ey,Machacek:1983fi},
 the matching conditions given in~\cite{Hambye:1996wb,Hempfling:1994ar,Holthausen:2011aa},
 i.e., matching of $\overline{\rm{MS}}$ coupling constants and pole masses to give boundary
 conditions of couplings up to two-loop order, are used. The top quark pole mass
 is chosen to be $M_{t}=173.3\pm2.8$ GeV~\cite{Alekhin:2012py},
 and other input parameters ($g_{1,2,3}$)
 are chosen as the central values~\cite{Beringer:1900zz}
 for most part of uncertainty is come from the top quark pole mass.
 The behavior of the Higgs mass with respect to scale $\mu$ is plotted
 in Fig.~\ref{fig:mHmu}. From the Left panel of the figure, we found that
 the 2-loop order corrections increases the $m_{H}(\mu)$ slightly compare with
 the 1-loop case, thus it indicates that it's proper to use Eq.~(\ref{eq:gammH2})
 when we study the RGE property of the Higgs mass at 2-loop order.
 And Fig.~\ref{fig:mHmu} depicts that the Higgs mass first growing and latter damping with
 the energy scale growing and is always larger than $zero$ at 1- and 2-loop level. In
 the right panel of Fig.~\ref{fig:mHmu}, the $M_{t}$ dependence of $m_{H}$ is considered.
 The scale at which
 the Higgs mass stops increasing and starts to decrease is around $10^{10}$ GeV for
 the case of $M_{t}=173.3$ GeV, and the scale decreases with the $M_{t}$ increasing.

\begin{figure}[!htp]
  \centering
  \includegraphics[width=0.45\linewidth]{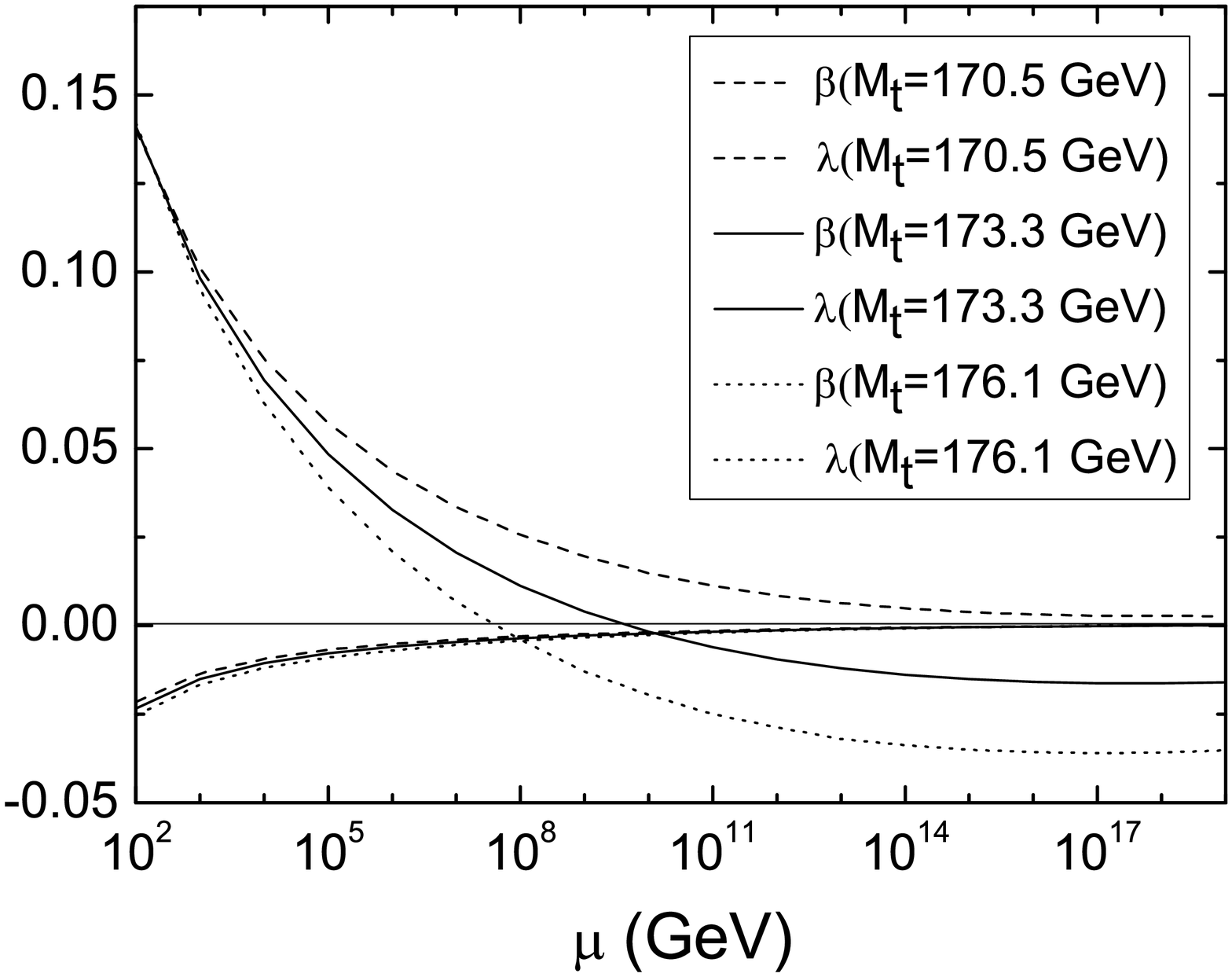}
  \centering
  \includegraphics[width=0.45\linewidth]{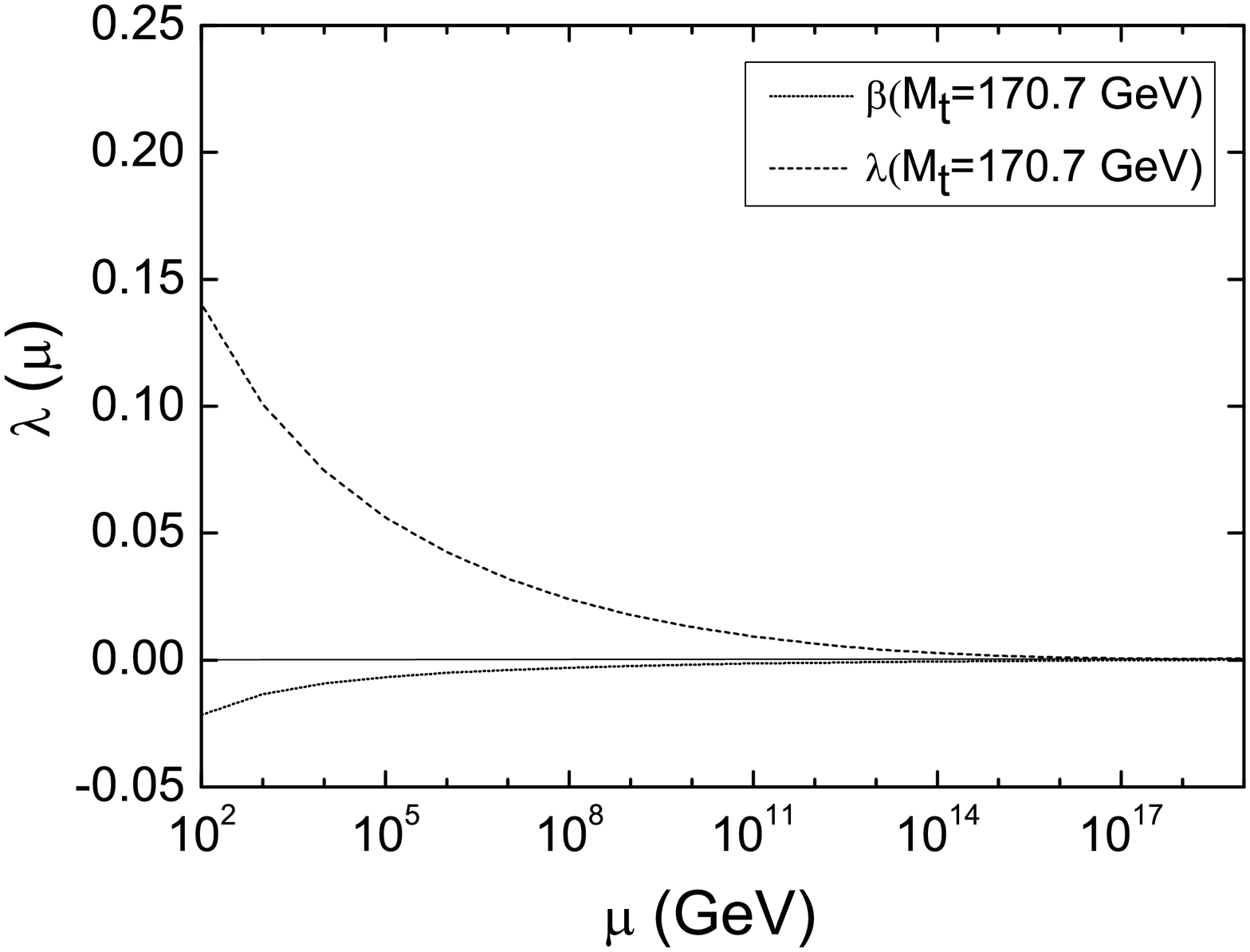}\\
  \caption{Left:Higgs quartic coupling/$\beta$-function behavior with/without considering RGE of $m_{H}$; Right:The behaviour of Higgs quartic coupling$\beta$-function for $M_{t}=170.7$ GeV.}\label{fig:lambda}
\end{figure}

The behaviours of Higgs quartic coupling/$\beta$-function with
and without considering the RGE of $m_{H}(\mu)$ are found to be the same
(plotted in Fig.~\ref{fig:lambda}) as it should be,
since the Higgs mass $m_{H}(\mu)$
do not enter into the the beta function of the Higgs quartic coupling explicitly
as that of other couplings.
From the figure, we found that
the $\lambda(\mu)$ damping to be negative
from about $\Lambda=$$10^{10}$ GeV for the central value of $M_{t}$.
And the scale
at which $\lambda(\mu)=0$ decreases with the value of $M_{t}$ decreasing,
as depicted on the left panel of Fig~.\ref{fig:lambda}.
The $\beta_{\lambda}$ approaches to
zero latter than $\lambda$, at about $10^{17}$ GeV, in agreement with~\cite{Degrassi:2012ry}.
For the input top quark pole mass $M_{t}=170.7$ GeV,
both $\beta/\lambda(\mu)$ approaches to zero around the scale $10^{17}$ GeV before the planck scale.
The vacuum stability of the SM is
achieved based on vacuum stability argument for $M_{t}<170.7$ GeV~\cite{Lindner:1988ww,Degrassi:2012ry}.

\section{Discussions and conclusions}
\label{sec:Higgsmass}

When the Higgs field is far larger than the mass term  in the Higgs potential~\cite{Lindner:1988ww,Sher:1988mj,Degrassi:2012ry},
\begin{eqnarray}\label{eq:effh}
V(H_{c})=-\frac{m^{2}}{2}H_{c}^{2}+\frac{\lambda}{4}H_{c}^{4}\approx\frac{\lambda}{4}H_{c}^{4},
\end{eqnarray}
with subscript $c$ to denote the classic Higgs field, i.e., the VEV of the Higgs
field with out source~\cite{Coleman:1973jx}.
Thus after investigated the scale-dependent property of the Higgs quartic coupling up to
two-loop level as in the last section, we may arrive at the vacuum stability region,
i.e., $\Lambda\leq10^{10}$ GeV with central top quark pole mass. The three-loop case are not
considered in this paper for which will change the vacuum stability condition tinny~\cite{Zoller:2012cv}.
The reasonable of the approximation in Eq.~(\ref{eq:effh}) is reconfirmed in the above section,
from the Fig.~\ref{fig:mHmu} we find that the Higgs mass changes little with energy growing and
always lower than 1 TeV, since the mass parameter m is connect with the Higgs mass through the
Higgs mechanism, the mass parameter $m$ will not
change too much with the energy scale growing and won't be very large, thus $m^{2}\ll H^{2}_{c}$
at the large $H_{c}$ scenario, i.e., at the large energy scale.
Then, the stability may be character by the sign of
$\lambda(\mu)$ very well.

Based on the vacuum stability argument, the Higgs quartic
coupling $\lambda<0$ at the energy above $\mu\sim10^{10}$ GeV for central
top quark pole mass, where the Higgs
potential will not be stable.
In the derivation process of Eq.~(\ref{eq:rgem}),
the tadpole renormalization played one important role, which makes Eq.~(\ref{eq:rgem})
gauge invariant~\cite{Fleischer:1980ub}.
And in other notation, the RGE of VEV in this scenario can be given by
\begin{eqnarray}\label{eq:rgev}
\gamma_{v^2}=v^2(\gamma_{m^2}/m^2-\frac{\beta_\lambda}{\lambda})
\end{eqnarray}
through connecting RG functions of broken ($v\neq0$)and
unbroken(v=0) phase~\cite{Jegerlehner:1} by defining the relationship
between the mass parameter m and the Higgs quartic coupling $\lambda$:
$m^2=\lambda v^2$. The $1/\lambda$ term is induced from the
tadpole renormalization.
The VEV behaves as $v^2(\mu^{2})\sim(\mu^{2})^{-\beta_{\lambda}(\mu^{2})
/\lambda(\mu^{2})}$ in high energy region~\cite{Jegerlehner:2012kn}.
According to
the behaviours of $\beta_\lambda(\mu)(\lambda(\mu))$ plotted in Fig.~\ref{fig:lambda},
$v$ decreases to one infinite small value as an
essential singularity at about the scale at which $\lambda=0$, and then increase gradually
and is always larger than zero.
Thus above the scale at which $\lambda=0$, one can always expect
$m^2_{H}(\mu)=2\lambda(\mu)v^{2}(\mu)<0$. Which conflicts with the positive
property of the Higgs mass as shown in Fig.~\ref{fig:mHmu}. And this kind of conflict
exists for $M_{t}>170.7$ GeV.

To avoid the conflict between the nonnegative property
of the Higgs mass and the negative value at high energy scale as
predicted by vacuum stability for $M_{t}>170.7$ GeV, we need to consider
that the VEV have the same rescalling property as the Higgs field,
\footnote{ We want to point out that, the key of the Randall-Sundrum
model to explore the hierarchy problem with one extral demension is just
based on the assumption that the VEV, $v$ have the same rescalling property
as that of the Higgs field in Wilsonian sense. One can refer to derivation
and arguments of formulas (17-21)in the paper\cite{Randall:1999ee}.}
then the renormalization constant of VEV need to be $Z_{H}$, i.e.,
$v_0=Z^{1/2}_{H}v$ but $v_0=Z^{-1/2}_{1}Z^{1/2}_{H}Z^{1/2}_{0}v$, which indicates that
$Z_{1}=Z_{0}$ in the broken phase.
Thus the RGE of the VEV could be
obtained through the anomalous dimension of the Higgs field~\cite{Arason:1991ic}.
And we have achieved this point based on argument on Wilsonian RG
method~\cite{Bian:2013xra}. While in this case the renormalization of the Higgs mass
term no longer preserves the gauge invariant property~\cite{Bian:2013xra}.
The RGE of VEV derived through the anomalous dimension of
the Higgs field is different from the RGE of VEV derived from Eq.~(\ref{eq:rgev})
~\cite{Jegerlehner:1,Jegerlehner:2013cta},
due to these two RG procedures are different in the renormalization of VEV.

Otherwise, there is one possibility that the conflict discussed above disappears.
Which happens when the top quark pole mass $M_{t}$ inputted
is smaller than $170.7$ GeV. Then it's suitable to consider the tadpole renormalization
as explored in this paper, and the using of the renormalization procedure of the Higgs mass
in this paper is a
top priority for the gauge invariant property is preserved there.

\appendix
\section{Integration formula involved in divergences calculation }
\label{sec:MI}
Scalar and tensor integrals involved in one-loop calculations,
 \begin{eqnarray}
 \int\frac{d^{d}k}{(2\pi)^{d}}\frac{1}{(k^{2}-m^{2})^{2}}&=&
 \frac{i}{(4\pi)^{2}}\frac{1}{2-d/2},\nonumber\\
\int\frac{d^{d}k}{(2\pi)^{4}}\frac{k^{\mu}k^{\nu}}{(k^{2}-m^{2})^{2}}
%&=&-\frac{g^{\mu\nu}}{d}
%\frac{i}{(4\pi)^{d/2}}\frac{\Gamma(1-d/2)}{\Gamma(1)}(\frac{1}{m^{2}})^{1-d/2}\nonumber\\
 %&&+i\frac{g^{\mu\nu}}{d}\frac{m^{2}}{(4\pi)^{d/2}}\frac{\Gamma(2-d/2)}
 %{\Gamma(2)(m^{2})^{2-d/2}}\nonumber\\
&=&-\frac{g^{\mu\nu}}{2}\frac{i}{4\pi}\frac{1}{1-d/2}
+i\frac{2g^{\mu\nu}}{4}\frac{m^{2}}{(4\pi)^{2}}\frac{1}{2-d/2},\nonumber\\
\int\frac{d^{d}k}{(2\pi)^{d}}\frac{k^{\mu}k^{\nu}}{(k^{2}-m^{2})^{3}}
&=&\frac{ig^{\mu\nu}}{4(4\pi)^{2}}\frac{1}{2-d/2},\nonumber\\
\int\frac{d^{d}k}{(2\pi)^{d}}\frac{k^{4}}{(k^{2}-m^{2})^{3}}
&=&-\frac{i}{4\pi}\frac{1}{1-d/2}+i3\frac{m^{2}}{(4\pi)^{2}}\frac{1}{2-d/2},
\nonumber\\
\int\frac{d^{d}k}{(2\pi)^{d}}\frac{k^{\mu}k^{\sigma}k^{\nu}k^{\rho}}
{(k^{2}-m^{2})^{4}}&=&\frac{i}{24(4\pi)^{2}}(g^{\mu\sigma}g^{\nu\rho}+
g^{\mu\nu}g^{\sigma\rho}
+g^{\mu\rho}g^{\sigma\nu})\frac{1}{2-d/2}.
\end{eqnarray}

\end{document}